\documentclass{article}

\usepackage{arxiv}

\usepackage[utf8]{inputenc} 
\usepackage[T1]{fontenc}    
\usepackage{hyperref}       
\usepackage{url}            
\usepackage{booktabs}       
\usepackage{amsfonts}       
\usepackage{nicefrac}       
\usepackage{microtype}      
\usepackage{graphicx}
\usepackage{lipsum}

\title{Real and virtual direct photon measurements with ALICE}

\author{
  H. Sebastian Scheid\\
  for the ALICE Collaboration\\
  \texttt{s.scheid@cern.ch} \\
}

\begin{document}
\maketitle
\begin{abstract}
In this contribution the latest measurements of real and virtual photons in Pb--Pb collisions at $\sqrt{s_{\rm NN}} = 5.02$ TeV from the ALICE Collaboration are presented. The extracted photon spectra are compared to predictions from state-of-the-art models that explicitly include the whole evolution of the heavy-ion collision as well as measurements at lower energies.
Experimental developments that in the future will help to extend the range and procession of the measurements are introduced before concluding with an outlook to possible future measurements at the LHC.
\end{abstract}

\section{Why Photons?}
If we think about the creation of the little big bang in the laboratory by colliding heavy nuclei at ultra-relativistic velocities we can identify different stages of these collisions in which different physics is at work.
While most partons in the nuclei will just pass through each other and deposit energy in the collision zone, some will interact in  most violent initial collisions. These initial scatterings can be described in perturbative quantum-chromo dynamics (pQCD).
After this the system will start to build up a medium which will approach equilibrium and form a plasma of quarks and gluons at a thermal scale of some hundred MeV. This can be described by modelling a pre-equilibrium phase followed by hydrodynamic evolution.
With the passing of time, this system will expand and cool down. Confinement will force the quarks and gluons into hadrons. Shortly after, the inelastic interactions will cease, followed by elastic scatterings.
Finally the debris of the collision will be measured in the detector.
While hadronic observables will always be obscured by interactions in the later stages of the collision, electromagnetic probes carry their information about the early stages of the heavy-ion collision to the detector unscathed.
In addition, a time ordering of the transverse momentum ($p_{\rm T}$), and in the dielectron case also invariant mass ($m_{\rm ee}$), is expected. Direct photons, that are produced earlier in the collision, tend to have higher $p_{\rm T}$ or $m_{\rm ee}$.
In the picture of an equilibrated thermal source producing the photons, the spectrum should depend on the size and the temperature of the source. While the absolute yield depends on the space time volume of the source, the inverse slope of the spectrum is only sensitive to the temperature $T$. In the case of real photons the energy of the photon measured in the detector is blue shifted due to the expanding source. For the virtual direct photons such a blue shift is not present since the Lorentz invariant mass of the virtual photons is the observable. In the terminology this is reflected by measuring an effective temperature in the real direct photon measurements.

While the above arguments are very compelling, the measurement of real and virtual direct photons are notoriously challenging.
In both measurements the main sources of real photons and dielectrons are decays of scalar and pseudo-scalar mesons.
In the case of virtual photons an additional physical background stems from the semi-leptonic decays of correlated open heavy-flavour (HF) hadrons.
Since not all decay products but only the leptons are reconstructed, the mass distribution of the HF contributions is wide and with little structure.
In addition it is not a priori clear which of the electrons originate from the same hadronic decay, which induces a combinatorial background of fake pairs in the dielectron analysis.
This combinatorial background of fake pairs is estimated by pairing $\rm{e^{+}e^{+}}$ and $\rm{e^{-}e^{-}}$ which are then subtracted from all $\rm{e^{+}e^{-}}$ pairs.

\section{Results}
Discussed here are measurements of photons in the ALICE detector using the internal and external conversion, by identifying the electron candidates. In addition ALICE has the possibility to directly measure photons in the electromagnetic calorimeters, e.g. the photon spectrometer (PHOS).
The presented data from Pb--Pb collisions at a center-of-mass energy per nucleon pair $\sqrt{s}_{\rm{NN}} = 5.02$ TeV was recorded with a minimum bias trigger (2015), and an online enhancement of events in the 0--10\% and 30--50\% centrality classes (2018).

To extract the direct-photon signal in the dielectron measurement a parameterisation of the invariant mass $m_{\rm ee}$ spectrum with a three component function is performed: $ f_{\rm fit} = r \times f_{\rm dir} + (1 - r) \times f_{\rm LF} + f_{\rm HF}$.
Here $f$ denotes the contributions from direct photons (dir), as well as from the light-flavour (LF) and the heavy-flavour (HF) hadron decays. The free parameter $r$ reflects, the fraction of virtual direct to inclusive virtual photons which is equivalent to the same fraction for real photons for $m_{\rm ee} \rightarrow 0$. The HF contribution is fixed to the expected contribution. The range of the parametrisation is chosen to be always above the mass of the $\pi^{0}$-meson.
To calculate the yield of real photons from $r$, the inclusive photon $\gamma^{\rm inc}$ measurement is needed. This was performed with the photon conversion method (PCM).


\begin{figure}[ht]
    \centering
    \begin{minipage}[b]{0.48\textwidth}
    \includegraphics[width=0.9\textwidth]{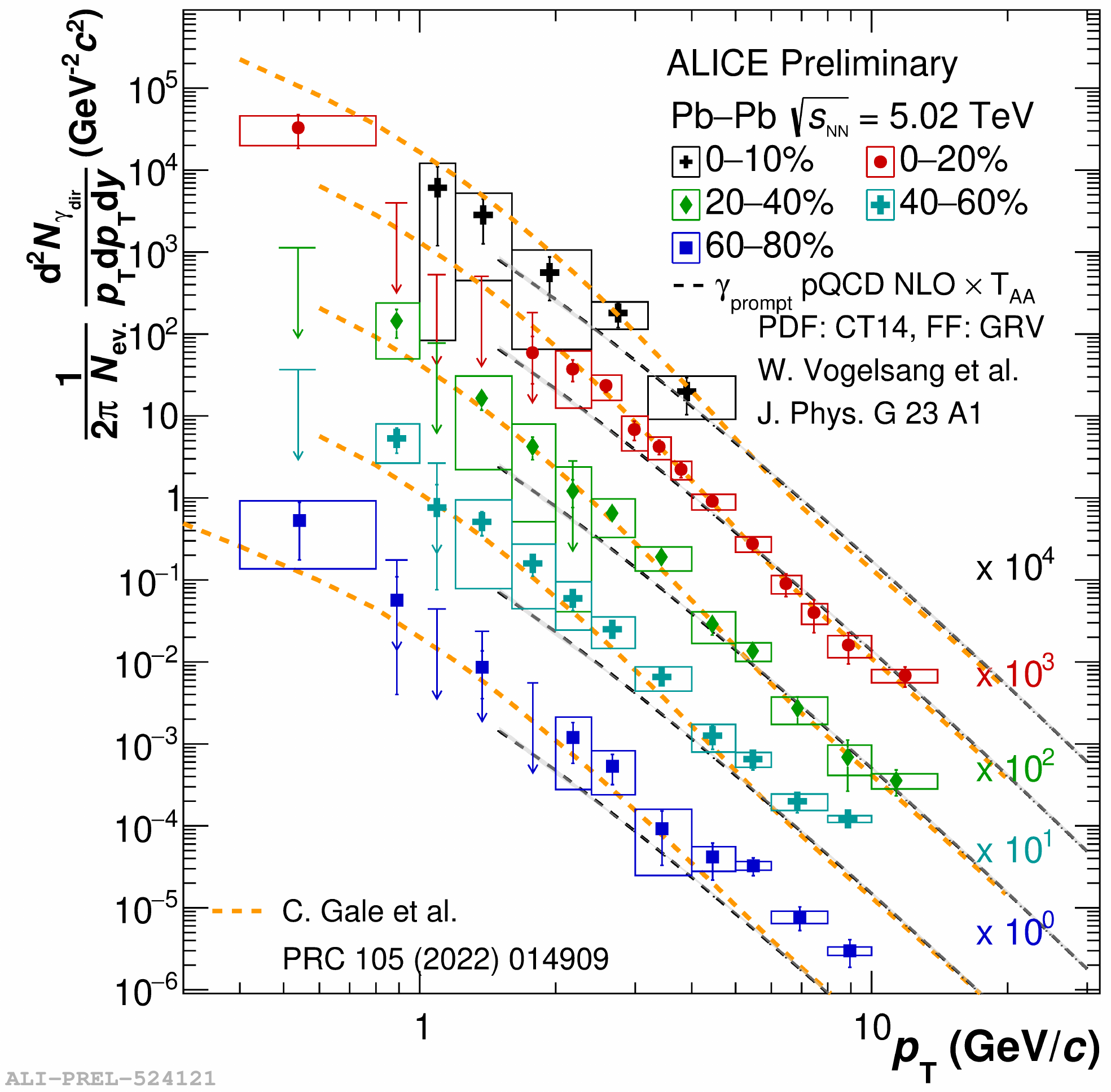}
    \caption{Invariant yield (left) of direct photons and $R_{\gamma}$ (right) at different centralities in Pb--Pb collisions at 5.02 TeV. The data are compared to pQCD~\cite{Vogelsang:1997cq} and model calculations that include all stages of the collision~\cite{Gale:2021emg}.}
    \end{minipage}
    \begin{minipage}[b]{0.48\textwidth}
    \includegraphics[width=0.9\textwidth]{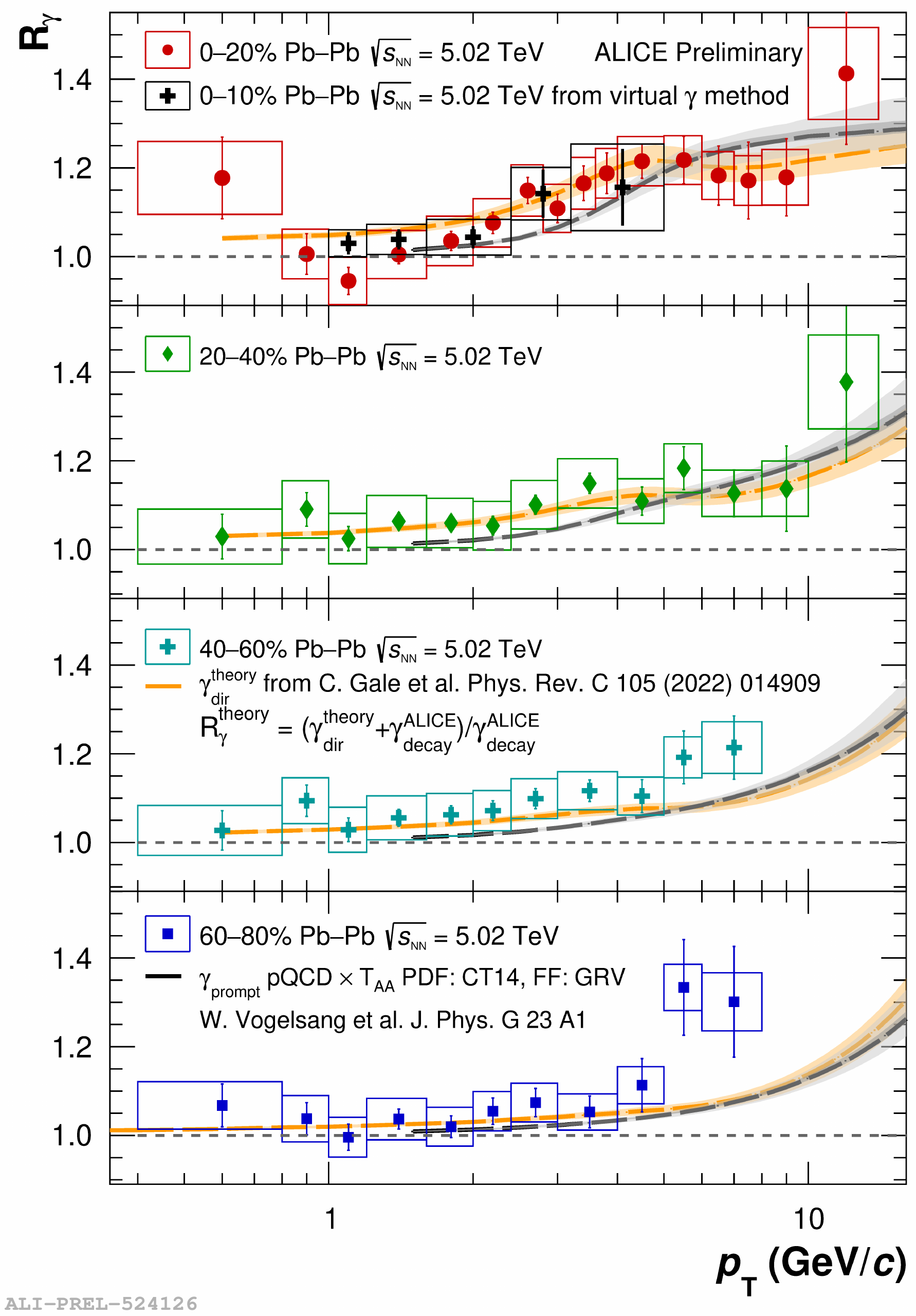}
    \end{minipage}

    \label{fig:directPhotons}
\end{figure}

The measurement of direct photons is shown for different centrality intervals in fig.~1 (left). The direct photons were measured  via reconstructing real photons in the PCM method and with the virtual direct photon method in the 0--10\% centrality class. Upper limits as a 95\% confidence interval (CL) are given in case of data points being consistent with 0 within statistical uncertainties. The data are compared to model calculations for direct photons from pQCD processes. In all centrality classes, the data tend to be above the pQCD prediction at an intermediate $p_{\rm T}$ range of 1-3 GeV/$c$ . A good agreement of the data in all measured centrality classes is achieved by model calculations that explicitly include all the different stages of a heavy-ion collision that emit photons, i.e. a pre-equilibrium as well as a QGP and hot hadronic phase~\cite{Gale:2021emg}.
The representation of the measurements in terms of the ratio of inclusive photons to decay photons $R_{\gamma} = \frac{\gamma^{\rm inc}}{\gamma^{\rm decay}}$ is shown in fig.~1 (right), for the different centrality classes. The $R_{\gamma}$ shows that at small $p_{\rm T}$ the signal of direct photons is expected to be in the order of 5\% wrt to the decay contribution for the given system. The measured data are in agreement with unity, but also the expected photon signal from model calculations~\cite{Gale:2021emg}. At larger $p_{\rm T}$ the data differs from unity indicating a significant source of photons that are not originating from hadronic decays. This signal is most pronounced in the most central collisions, and fades away when decreasing the centrality. This trend is well described by the model calculations.

\begin{figure}[ht]
    \centering
    \begin{minipage}{0.52\textwidth}
    \includegraphics[width=1.0\textwidth]{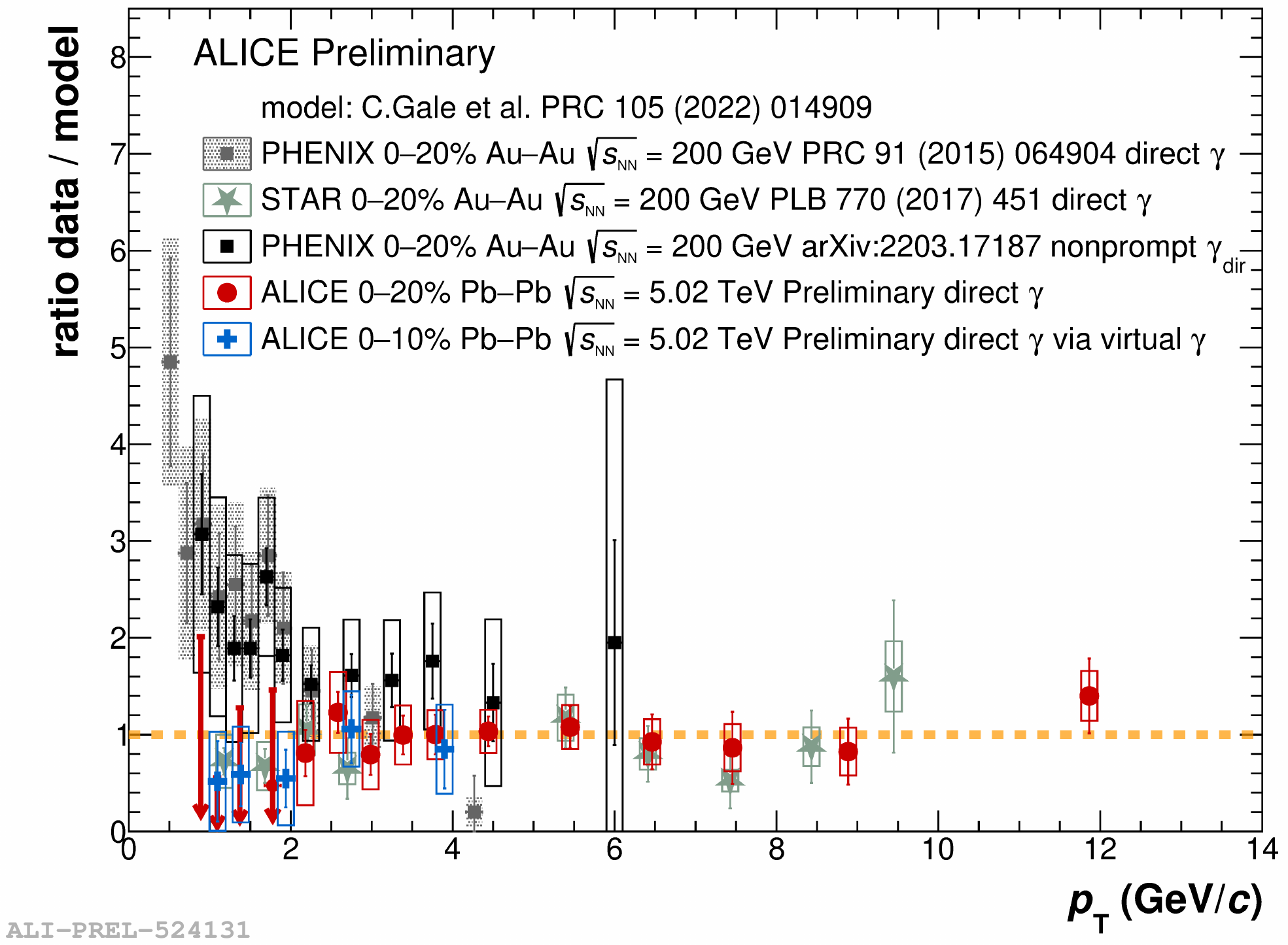}
    \end{minipage}
    \begin{minipage}{0.47\textwidth}
    \includegraphics[width=0.85\textwidth]{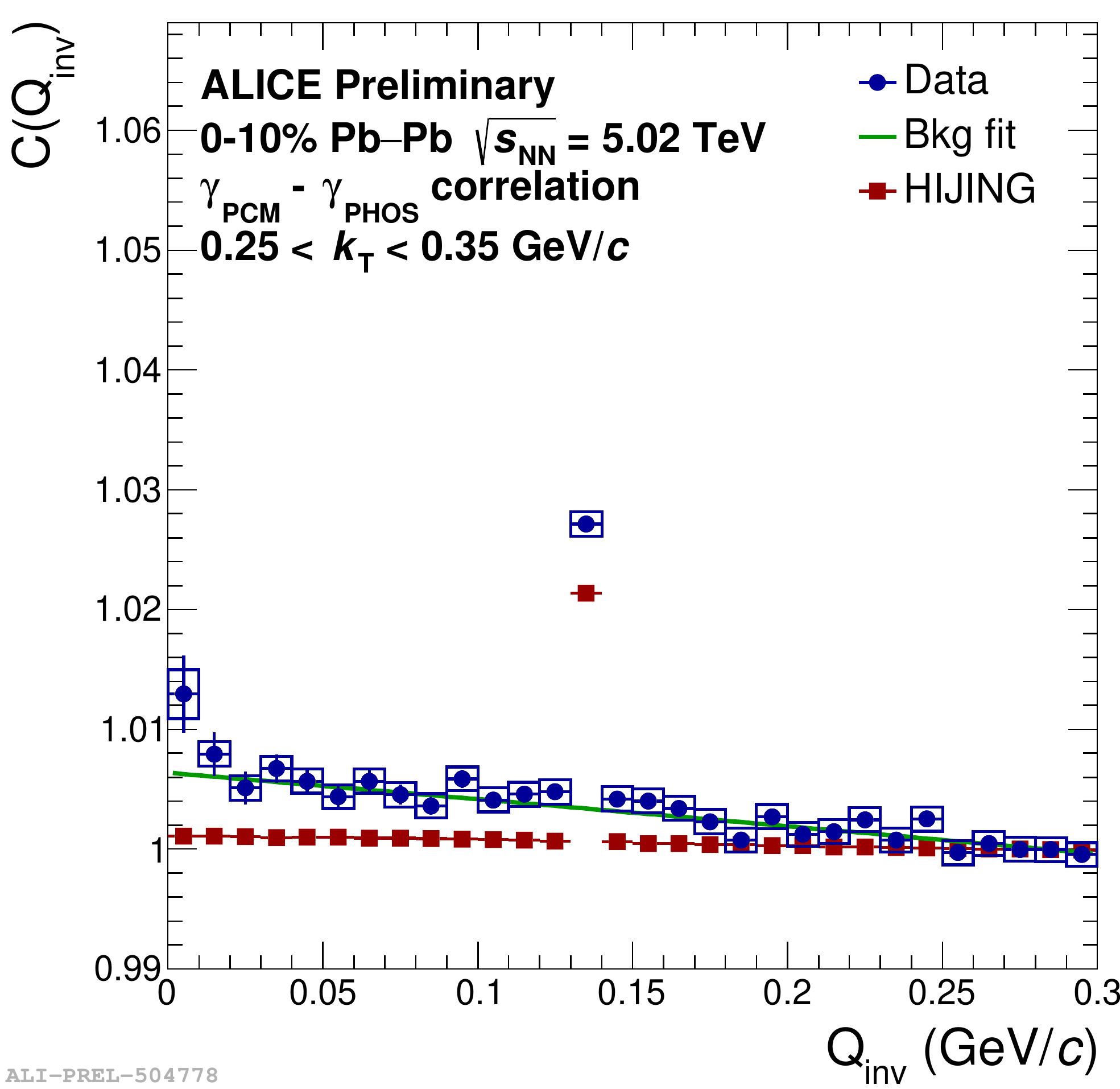}
    \end{minipage}
    \caption{Measured direct photon yields from the STAR, the PHENIX, and ALICE collaborations are compared in terms of their ratio to a state-of-the-art model at their respective energies (left). Correlation function $C$ of two photons as a function of $\rm Q_{inv}$ in the range of $0.25 < k_{\rm T} < 0.35$ GeV/$c$ in data and a HIJING simulation in 0--10\% Pb--Pb collisions (right).}
    \label{fig:comparison}
\end{figure}

A comparison of the ALICE measurements with previous measurements of direct photons by the STAR and the PHENIX collaborations is shown in fig.~\ref{fig:comparison}. To compare the data at different collision energies and centralities, the ratios to the same model~\cite{Gale:2021emg} are compared.
While the data from the virtual and real photons from ALICE and the real photon measurement by the STAR collaboration are in agreement with each other and the model calculations, the two measurements by the PHENIX collaboration are showing a slight tension wrt the model calculations.
However, taking the systematic uncertainty that dominate both measurements serious, none of the measurements shows a tension with this state-of-the-art model.

An alternative approach to extract the direct-photon signal is through Hanbury-Brown-Twiss (HBT) correlations. Following the WA98 collaboration~\cite{WA98:2003ukc} the correlation function $C$ of two photons from the same source as a function of the photon pair relative momentum $Q_{inv}$ is calculated as: ${\rm C(Q_{inv}) = Q_{inv,SE}/Q_{inv,ME}}$. Where SE and ME denote pairs from the same and different (mixed) events, respectively.
Figure~\ref{fig:comparison} (right) shows the ALICE result in the two photon average momentum interval $0.25 < k_{\rm T} < 0.35$ GeV/$c$ for data and HIJING simulation in 0--10\% most central Pb--Pb collisions. In this analysis one photon was reconstructed with the PCM while the second was measured with the PHOS. The sharp peak in simulation and data is expected from the $\pi^{0} \rightarrow \gamma\gamma$ decay. In the case of photons, ie vanishing mass, $\rm Q_{inv}$ reduces to the invariant mass of the photon pair $m_{\gamma\gamma}$. A residual background in the data from particle showers is parametrised with a second order polynomial. At small $\rm Q_{inv}$ a hint for an additional signal is quantified by fitting the data with eq.~\ref{eq:cfit}, including the correlation strength $\rm \lambda_{inv}$ and the source size $\rm R_{inv}$.

\begin{equation}
\rm C(Q_{inv}) = 1+ \lambda_{inv} \exp (-R^{2}_{inv}Q_{inv}^2)
\label{eq:cfit}
\end{equation}

The large statistical and systematic uncertainties unfortunately do not allow for a conclusion, but the extraction of the correlation function could in the future be a good tool to distinguish early and late emission of photons~\cite{Garcia-Montero:2019kjk}.

\begin{figure}[ht]
    \centering
    \begin{minipage}{0.48\textwidth}
        \includegraphics[width = 0.9\textwidth]{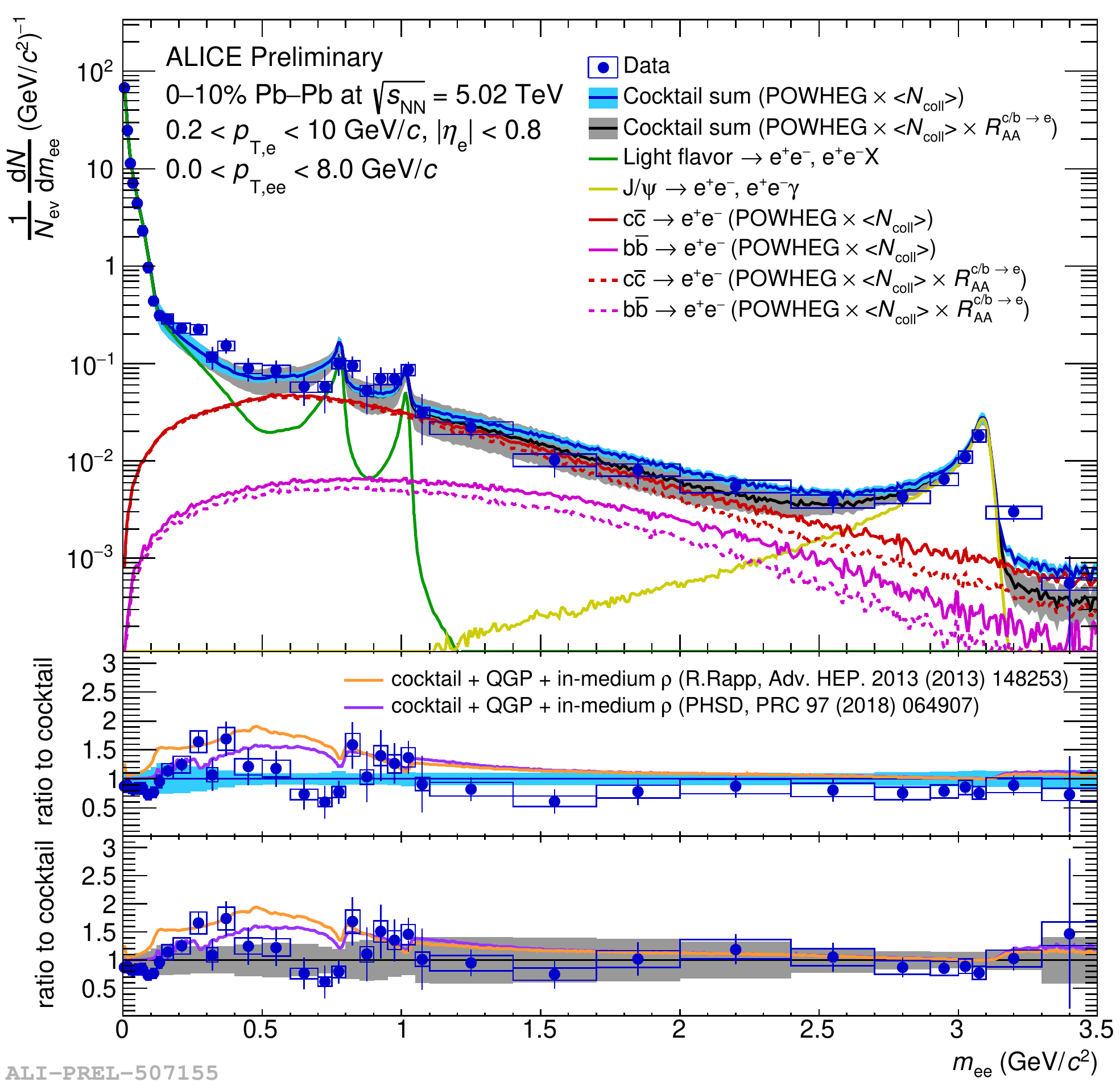}

    \end{minipage}
    \begin{minipage}{0.48\textwidth}
        \includegraphics[width = 0.9\textwidth]{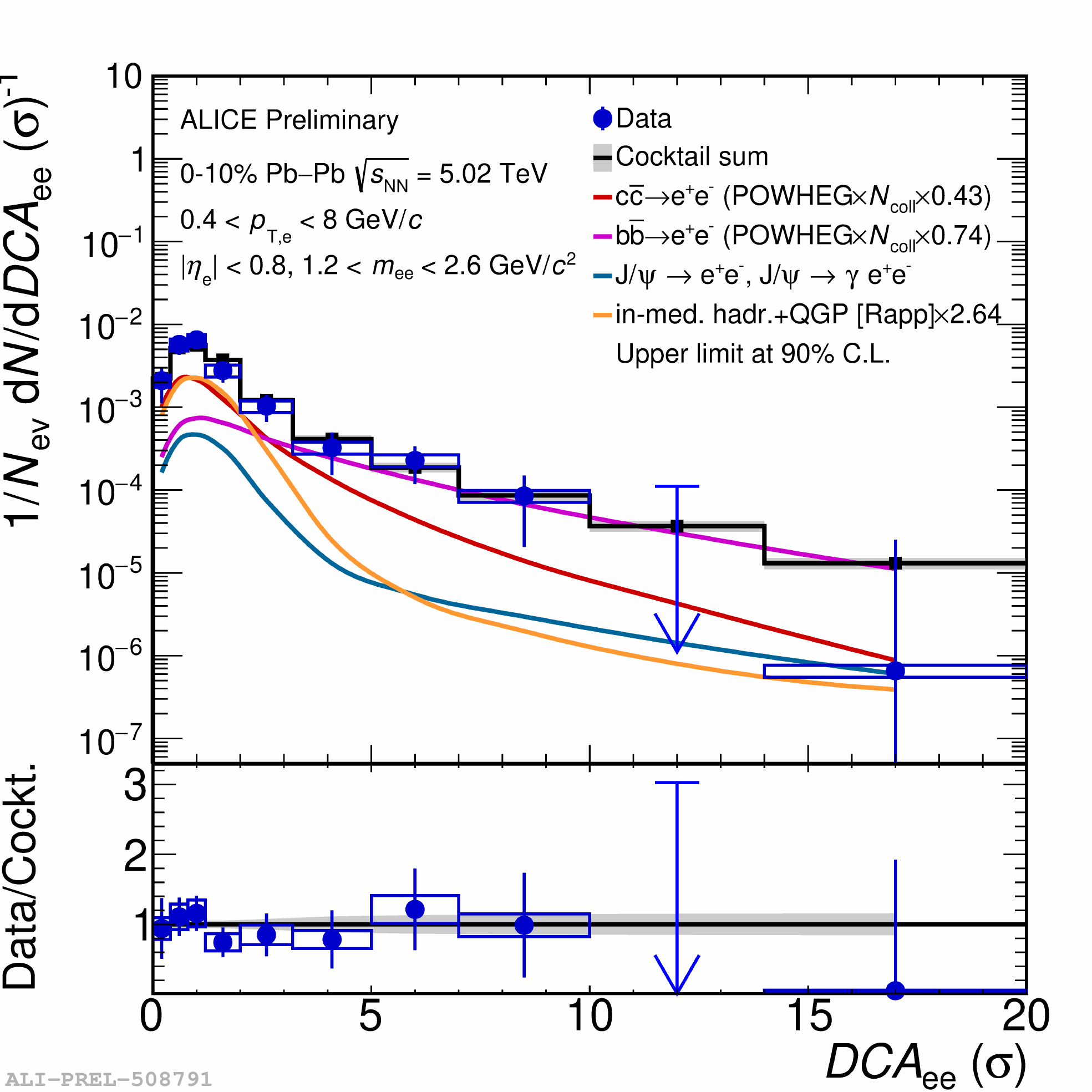}

    \end{minipage}
    \caption{Left: Dielectron yield as a function of $m_{\rm ee}$ measured in 0--10\% Pb--Pb collisions at $\sqrt{s_{\rm NN}}=5.02$~TeV (top panel). The data is compared to hadronic cocktail simulations. In one case (solid lines) the heavy-flavour contribution is based on the measured cross section in pp collisions~\cite{ALICE:2018fvj} and scaled with $N_{\rm coll}$. In the second case (dashed lines) the measured $R_{\rm AA}$ of electrons from heavy-flavour hadron decays is included as described in the text. The respective ratios of the data to the cocktails are shown in the middle and bottom panel. In both cases also predictions for thermal dielectrons from a hadronic and partonic phase are indicated~\cite{Song:2018xca,Rapp:2013nxa}.
    Right: Dielectron yield as a function of $\rm DCA_{ee}$ in the intermediate mass region. If the data are consistent with zero upper limits are given. The spectrum was parametrised with templates for charm, beauty and a prompt contribution.}
    \label{fig:pbpbDielectrons}
\end{figure}

One of the motivations given in the beginning of this article is the measurement of the average temperature in the early phase of the collision by extracting the inverse slope of the virtual direct-photon invariant mass spectrum at masses above the $\phi$ meson. Figure~\ref{fig:pbpbDielectrons} shows the measured dielectron yield in the ALICE acceptance as a function of the invariant mass. The top panel shows the measurement compared with the expectation from hadronic decays, the so-called hadronic cocktail. Two different approaches for the contribution from heavy-flavour hadron decays are presented. In the first case the measured cross section of dielectrons from heavy-flavour hadron decays measured in pp collisions at $\sqrt{s} = 5.02$ TeV~\cite{ALICE:2020mfy} is scaled with the number of binary nucleon-nucleon collisions ($N_{\rm coll}$) (solid lines). The functional shape as a function of mass is based on the event generator POWHEG. Measurements of the nuclear modification factor $R_{\rm AA}$ of electrons from charm and beauty quarks however show a significant modification, of the $p_{\rm T}$ spectra~\cite{ALICE:2019nuy}.
To successfully describe the observed modification models need to include initial cold nuclear matter (CNM) effects as well as later stage energy loss.
The problem in propagating this measured modification to the dielectron spectra is that the modifications from CNM and energy loss effects affect the dielectron spectra in different ways. While the energy loss affects each of the quarks, and thus the electrons, independently, the CNM effects affect the pair. To accommodate for this an approach was chosen that is able to include both effects with the expected impact on the pair.
As a starting point the measured $R_{\rm AA}$ of $\rm c,b \rightarrow e$ was parameterised. To disentangle CNM and energy loss (EL) effects a calculation based on the EPS09 nPDF~\cite{Eskola:2009uj} set was used. By dividing the data parameterisation by the nPDF expectation, one can construct the expectation for the $R_{\rm AA}$ that originates only from the energy loss of the heavy quark.
In a next step two different $R_{\rm AA}$ for the dielectrons, one for energy loss and one for CNM effects, can be calculated in a Monte-Carlo approach. Here the single electron $R_{\rm AA}$ are used as weights. In the energy loss case factorisation is assumed and the weight for the pair is calculated as the product of the electron weights $R^{\rm HFee}_{\rm AA, EL} = w_{1} \times w_{2}$. In the case of CNM effects no factorisation is expected and the weight for the pair is the average of the electron weights $R^{\rm HFee}_{\rm AA, CNM} = (w_{1}+w_{2})/2$.
The heavy-flavour dielectron nuclear modification factor can then be calculated as $R^{\rm HFee}_{\rm AA} = R^{\rm HFee}_{\rm AA,CNM} \times R^{\rm HFee}_{\rm AA,EL}$. This factor is then used to modify the cocktail based on $N_{\rm coll}$-scaling. The effect can be seen when comparing the dashed lines in the upper left panel of fig.~\ref{fig:pbpbDielectrons} with the solid lines.
The middle and bottom panels show the ratio of the measurement to the cocktail expectation for the case of $N_{\rm coll}$ scaling and including the $R^{\rm HFee}_{\rm AA}$, respectively. In the mass region between the $\phi$ and the $J/\psi$ we can see that the data is below the expectation from $N_{\rm coll}$-scaling (middle panel) which is expected, since there is an overall suppression seen in the $R^{\rm c,b \rightarrow e}_{\rm AA}$. Including the $R^{\rm HFee}_{\rm AA}$ estimate slightly brings down the cocktail which leads to the cocktail being closer to the measurement (bottom panel). However, as discussed before, the uncertainties are large, and no conclusion is possible. Calculations for a dielectron contribution from a hadronic and a partonic phase~\cite{Rapp:2013nxa,Song:2018xca} are shown in both ratios to indicate the size of the expected signal.


A different and very promising approach is to separate the dielectrons from heavy-flavour hadron decays from the prompt thermal contribution based on their decay topology. The basic idea is simple: the finite decay length of charm and beauty hadrons can be exploited to separate the contributions of heavy-flavour and prompt dielectrons. For this the pair distance-of-closest-approach ($\rm DCA_{ee}$) is constructed as:

\begin{equation}
    \rm DCA_{ee} = \sqrt{\frac{DCA_{1}^{2}+DCA_{2}^{2}}{2}}.
\end{equation}

$\rm DCA_{1}$ and $\rm DCA_{2}$ denote the DCA of the tracks to the reconstructed primary vertex, normalised to the respective resolution. By construction we expect a ordering of the distributions depending on the decay length ($c\tau$) of the decaying particle: $\rm DCA_{ee}(prompt) < DCA_{ee}(charm) < DCA_{ee}(beauty)$~\cite{ALICE:2018fvj}.
A big advantage of this approach is its model agnostics. Figure~3 (right) shows the data in the mass interval of 1.2 - 2.6 GeV/$c^{2}$.
The data was parametrised with templates for charm, beauty, $J/\psi$, and a prompt contribution that can be interpreted as a thermal source. While the uncertainties are large, the data are well described when including the thermal contribution dominant at low $\rm DCA_{ee}$, and suppression of the heavy-flavour contributions.



\section{Conclusion and outlook}

The measurement of the direct-photon yield in Pb--Pb collisions at $\sqrt{s_{\rm NN}} = 5.02$~TeV in several centrality classes by the ALICE Collaboration are well described by model calculations when including all stages of the heavy-ion collision. The comparison of measurements to those from the STAR and PHENIX collaborations with respect to the model predictions at the respective energies shows a good agreement of the ALICE and STAR measurements with the model. At low $p_{\rm T}$ the central values of the PHENIX results tend to be above the model expectation, however with large uncertainties. An alternative approach to access the direct-photon contribution is the measurement of the correlation function of the direct photons. A proof of concept was presented, but better statistical precision will be needed to conclude.

To access the average temperature in the earlier stages of the heavy-ion collision, i.e. the QGP, it is necessary to understand the dielectron contribution from charmed hadron decays including its modification. An approach to include these was discussed, however suffers from large uncertainties due to its model dependences.
A second, model agnostic, approach uses the decay typologies of the respective contributions. Based on the decay length of the mother particle the charm and beauty contributions can be singled out and a possible thermal source can be isolated. Both approaches have their advantages and disadvantages, while the modeling in the cocktail introduces model dependencies and large uncertainties, the topological separation however requires a large amount of statistics.

The recent upgrades of the ALICE TPC and ITS grant a better pointing resolution (factor 3-6) and increased readout rate (factor 100). Both greatly benefit the measurements of real and virtual photons.
For the 2030s the ALICE collaboration foresees the construction of a next-generation heavy-ion experiment~\cite{ALICE:2803563}. The ALICE 3 detector concept is based around a high-acceptance all-silicone tracker with unprecedented precision, complemented by particle identification. Its possibility to efficiently reject HF contributions to the dielectron spectrum together with the boost in effective statistics will enable multi-differential analyses. These types of measurements are expected to be sensitive to the time evolution of the QGP, and its early phases.

\bibliographystyle{plain}
\bibliography{literature}

\begin{thebibliography}{10}

\bibitem{ALICE:2018fvj}
Shreyasi Acharya et~al.
\newblock {Dielectron production in proton-proton collisions at $ \sqrt{s}=7 $
  TeV}.
\newblock {\em JHEP}, 09:064, 2018.

\bibitem{ALICE:2020mfy}
Shreyasi Acharya et~al.
\newblock {Dielectron production in proton-proton and proton-lead collisions at
  $\sqrt{s_{NN}}=$ 5.02 TeV}.
\newblock {\em Phys. Rev. C}, 102(5):055204, 2020.

\bibitem{ALICE:2019nuy}
Shreyasi Acharya et~al.
\newblock {Measurement of electrons from semileptonic heavy-flavour hadron
  decays at midrapidity in pp and Pb-Pb collisions at $\sqrt{s_{\rm{NN}}}$ =
  5.02 TeV}.
\newblock {\em Phys. Lett. B}, 804:135377, 2020.

\bibitem{WA98:2003ukc}
M.~M. Aggarwal et~al.
\newblock {Interferometry of direct photons in central Pb-208+Pb-208 collisions
  at 158-A-GeV}.
\newblock {\em Phys. Rev. Lett.}, 93:022301, 2004.

\bibitem{ALICE:2803563}
ALICE.
\newblock {Letter of intent for ALICE 3: A next generation heavy-ion experiment
  at the LHC}.
\newblock Technical report, CERN, Geneva, 2022.

\bibitem{Eskola:2009uj}
K.~J. Eskola, H.~Paukkunen, and C.~A. Salgado.
\newblock {EPS09: A New Generation of NLO and LO Nuclear Parton Distribution
  Functions}.
\newblock {\em JHEP}, 04:065, 2009.

\bibitem{Gale:2021emg}
Charles Gale, Jean-Fran\c{c}ois Paquet, Bj\"orn Schenke, and Chun Shen.
\newblock {Multimessenger heavy-ion collision physics}.
\newblock {\em Phys. Rev. C}, 105(1):014909, 2022.

\bibitem{Garcia-Montero:2019kjk}
Oscar Garcia-Montero, Nicole L\"oher, Aleksas Mazeliauskas, J\"urgen Berges,
  and Klaus Reygers.
\newblock {Probing the evolution of heavy-ion collisions using direct photon
  interferometry}.
\newblock {\em Phys. Rev. C}, 102(2):024915, 2020.

\bibitem{Rapp:2013nxa}
Ralf Rapp.
\newblock {Dilepton Spectroscopy of QCD Matter at Collider Energies}.
\newblock {\em Adv. High Energy Phys.}, 2013:148253, 2013.

\bibitem{Song:2018xca}
Taesoo Song, Wolfgang Cassing, Pierre Moreau, and Elena Bratkovskaya.
\newblock {Open charm and dileptons from relativistic heavy-ion collisions}.
\newblock {\em Phys. Rev. C}, 97(6):064907, 2018.

\bibitem{Vogelsang:1997cq}
W.~Vogelsang and M.~R. Whalley.
\newblock {A Compilation of data on single and double prompt photon production
  in hadron hadron interactions}.
\newblock {\em J. Phys. G}, 23:A1--A69, 1997.

\end{thebibliography}

\end{document}